\documentclass[letterpaper, 10 pt, conference]{ieeeconf}
\IEEEoverridecommandlockouts                              
\usepackage[dvipsnames]{xcolor}
\usepackage{tikz}
\newcommand*{\orangecircle}{\protect \tikz
  \protect \draw[orange,fill=orange] (0,0) circle (.5ex); }

\usepackage{algorithmic}
\usepackage{amsmath,amssymb,latexsym}
\usepackage{graphicx}
\usepackage{hyperref}
\allowdisplaybreaks

\newtheorem{definition}{Definition}
\newtheorem{lemma}{Lemma}
\newtheorem{remark}{Remark}

\title{\LARGE \bf Neural Energy Casimir Control for Port-Hamiltonian Systems}

\usepackage{times}
\usepackage{color, soul,bm}

\author{Liang Xu, Muhammad Zakwan and Giancarlo Ferrari-Trecate 
\thanks{This research is supported by the Swiss National Science Foundation under the NCCR Automation (grant agreement 51NF40 180545).}
\thanks{Authors are with the Institute of Mechanical Engineering, Ecole Polytechnique Fédérale de Lausanne (EPFL), CH-1015 Lausanne, Switzerland {\tt\small \{liang.xu, muhammad.zakwan, giancarlo.ferraritrecate\}@epfl.ch }}%
}

\begin{document}

\maketitle
\thispagestyle{empty}
\pagestyle{empty}

\begin{abstract}%
    The energy Casimir method is an effective controller design approach to stabilize port-Hamiltonian systems at a desired equilibrium.
    However, its application relies on the availability of suitable Casimir and Lyapunov functions, whose computation are generally intractable.
    In this paper, we propose a neural network-based framework to learn these functions.    
    We show how to achieve equilibrium assignment by adding suitable regularization terms in the training cost.
  We also propose a parameterization of Casimir functions for reducing the training complexity.
  Moreover, the distance between the equilibrium of the learned Lyapunov function and the desired equilibrium is analyzed, which indicates that for small suboptimality gaps, the distance decreases linearly with respect to the training loss.
  Our methods are backed up by simulations on a pendulum system.
\end{abstract}

\section{Introduction}
Port-Hamiltonian systems (PHSs) describe real-world applications as the interconnection of dynamical energy-storing, static energy-dissipating, and static lossless energy-routing elements. 
The coupling among these elements is realized through pairs of variables called ports.
Since energy is the building block among various physical domains, port-Hamiltonian modeling provides a generic framework for describing physical systems, and has been extensively used in various areas, for example, mechanical~\cite{van2014port, schaller2021control, tsolakis2021distributed} and electrical systems~\cite{strehle2020scalable, padilla9147748}.
The increasing significance of the port-Hamiltonian framework for modeling, control, and analysis of multi-physics systems is also evident from several monographs, e.g.~\cite{brogliato2020dissipative, vanderSchaft2017}.

PHSs are naturally passive, i.e., the increase in the system energy (described by a scalar function, called Hamiltonian), is smaller than the supply rate (the product of the input and the output ports).
Passivity implies that the Hamiltonian may serve as a Lyapunov function, and negative output feedback can asymptotically stabilize the PHS to the local minimum of the Hamiltonian.
However, if the Hamiltonian does not have a minimum at the desired equilibrium, the Hamiltonian  itself alone is not a Lyapunov function and negative output feedback control fails.
A well-known method for overcoming this obstacle is the energy Casimir method,
which uses additional conserved quantities (called Casimir functions, or Casimirs) in addition to the Hamiltonian for constructing the desired Lyapunov function~\cite{ortega1999energy, ortega2008control}.
Specifically, this approach interconnects the plant PHS with a controller in the port-Hamiltonian form and aims to generate Casimir functions, such that the closed-loop system has a well-defined Lyapunov function with a minimum located at the desired equilibrium.
Then, an additional damping term is employed to asymptotically stabilize the closed-loop system.

Unfortunately, the computation of Casimirs and Lyapunov functions requires to solve a set of partial differential equations (PDEs), which usually are analytically intractable and computationally challenging \cite{wu2020stabilization}.
Motivated by the universal approximation capability of neural networks (NNs) and available open-source machine learning frameworks for training NNs, in this paper, we propose a NN-based approach to learn Casimirs and Lyapunov functions in the energy Casimir method and shape the equilibrium of the closed-loop system without solving convoluted PDEs.
More in general, there is a also growing interest in using NNs for designing controllers or certifying system properties.
For example, \cite{chang2020neural} and \cite{dai2021lyapunov} propose iterative procedures to construct a neural controller and a neural Lyapunov function with provable stability guarantees.
\cite{yin2021stability} and \cite{pauli2021linear} abstract NNs using integral quadratic constraints and analyze the stability of feedback systems with neural controllers.
Graph NNs are used for distributed control in~\cite{yang2021communication} and \cite{gama2021graph}, and sufficient conditions on input-state stability are further provided in~\cite{gama2021graph}.
\cite{RN11575} and~\cite{chen2021learning} use NNs to learn the region of attraction of nonlinear systems for obtaining stability and safety certificates.
The above works mainly deal with either linear systems or very general classes of non-linear systems.

In this paper, we focus on port-Hamiltonian models and use NNs to tackle the computational obstacles in the application of the energy Casimir method.
The proposed framework is much simpler and computational less expensive than the methods in \cite{chang2020neural, dai2021lyapunov}, which deal with general nonlinear systems.
We show how to transform the Casimir design problem into a parametric optimization problem by using NNs to approximate the controller Hamiltonian, the Casimirs and the Lyapunov function for the closed-loop system.
To assign the minimum of the Lyapunov function to the desired equilibrium point, we use a novel training cost that penalizes the norm of the Jacobian and the eigenvalues of the Hessian of the neural Lyapunov function evaluated at the desired equilibrium point.
Moreover, we incorporate PDE constraints on Casimirs in the cost as a regularization term.
However, embedding PDE constraints in the cost might significantly increase the computational complexity of the training process.
Therefore, we further present a NN-based parameterization of Casimirs to satisfy the PDE constraints by design, which achieves a trade-off between the representation generality and the training speed.
Moreover, we provide an upper bound on the difference between the desired and achieved equilibrium in terms of the training loss, which shows that for small suboptimality gaps, the equilibrium assignment error scales linearly with the training loss.
As a benchmark, we consider set-point control of a pendulum and show that our method asymptotically stabilizes the pendulum to the desired equilibrium and learns a neural Lyapunov function with a large region of attraction.
A related work is~\cite{RN11663}, where the authors propose a stable NN-based controller structure based on the energy shaping control of Lagrangian systems.
Compared with~\cite{RN11663}, this paper deals with Hamiltonian systems and focuses on improving the classic energy Casimir method.

{\bf Organization}. The paper is organized as follows.
Section~\ref{sec.preliminary} recalls preliminaries on PHSs and the classic energy Casimir method.
Section~\ref{sec.NeuralCasimirControl} presents our main results on the parameterization of  Casimirs and a novel optimization cost for achieving the equilibrium assignment.
Section~\ref{sec.Simulation} discusses simulation experiments. Concluding remarks are given in Section~\ref{sec.Conclusions}.     

{\bf Notations}. For a differentiable function $H: \mathbb{R}^n\rightarrow \mathbb{R}$, $\frac{\partial H}{\partial \bm{x}}(\bm{x})$ denotes the column vector of partial derivatives of $H$. $\frac{\partial^\top H}{\partial \bm{x}}(\bm{x})$ is the transpose of $\frac{\partial H}{\partial \bm{x}}(\bm{x})$.

\section{Preliminaries\label{sec.preliminary}}
In this section, we briefly introduce PHSs and the energy Casimir control method.
\subsection{Port-Hamiltonian System}
An input-state-output PHS can be expressed as
\begin{equation}
\begin{aligned}
    \dot{\bm{x}}& = (\bm{J}(\bm{x})-\bm{R}(\bm{x}))   \displaystyle{ \frac{\partial {H}(\bm{x})}{\partial \bm{x}}}+\bm{G}(\bm{x})\bm{u}, \\
    \bm{y}& = \bm{G}^\top(\bm{x})  \displaystyle{\frac{\partial {H}(\bm{x})}{\partial \bm{x}}}\, , 
\end{aligned}\label{eq.Plant}
\end{equation}
where $\bm{x} \in \mathbb{R}^n$ is the state, $\bm{u}\in \mathbb{R}^m$ is the input, and $\bm{y} \in \mathbb{R}^m$ is the measured output;
$\bm{J}(x)$, $\bm{R}(x)$ are the interconnection and damping matrices, respectively, satisfying $\bm{J}(\bm{x}) = -\bm{J}^\top (\bm{x})$ and $\bm{R}(\bm{x}) = \bm{R}^\top (\bm{x}) \geq 0$;
$\bm{G}(\bm{x})$ is the input matrix, assumed to be full rank;
$H: \mathbb{R}^n \rightarrow \mathbb{R}$  is a continuously differentiable function, called the \emph{Hamiltonian} of the system.
By construction, one has the passivity property
\begin{equation}
\label{pass_cond}
    \dot{{H}}(\bm{x})\leq \bm{y}^\top \bm{u}\, ,
\end{equation}
which is extensively used in the stability analysis of PHSs~\cite{vanderSchaft2017}.

\subsection{Casimir Function}

Casimirs are conserved quantities along the state trajectories of a PHS.
\begin{definition}[Casimir function]
A continuously differentiable function $C(\bm{x}): \mathbb{R}^n \rightarrow \mathbb{R}$ is a Casimir function for~\eqref{eq.Plant}, if it satisfies
\begin{align}
  \label{eq.CasimirCondition}
\frac{\partial^\top C}{\partial \bm{x}}(\bm{x}) (\bm{J}(\bm{x}) -\bm{R}(\bm{x}))=0 , \quad \bm{x} \in \mathbb{R}^n.
 \end{align}
\end{definition}


 From the above definition, we know that adding a Casimir to the Hamiltonian function will not change the dynamics~\eqref{eq.Plant}.
 Therefore, Casimirs can be used to modify the Hamiltonian of the system, which  is the motivation for the energy Casimir method introduced below.

\subsection{Energy Casimir Method}
We are interested in set-point control of~\eqref{eq.Plant}, that is in designing a controller to stabilize~\eqref{eq.Plant} to a given set-point $\bm{x}^*$.
If  $\bm{x}^*$  is a local minimum of the Hamiltonian function $H$, due to the passivity property~\eqref{pass_cond}, a negative output feedback  controller, i.e., $\bm{u}=-\bm{y}$, would achieve the goal~\cite{vanderSchaft2017}.
However, if $\bm{x}^*$ is not a local minimum of $H$, this approach can not be applied and alternative methods are needed.

The energy Casimir method tackles this issue by introducing a controller in the PHS form and designing an appropriate Casimir and a Lyapunov function to place the closed-loop system minimum at $(\bm{x}^*, \bm{\xi}^*)$ for a suitable controller equilibrium point  $\bm{\xi}^*$ of the controller state. 
The design procedure is thoroughly described in~\cite{vanderSchaft2017} and summarized below for convenience.
Consider the PHS controller
\begin{equation}
\begin{aligned}
    \dot{\bm{\xi}}& =\left[\bm{J}_{c}(\bm{\xi})-\bm{R}_{c}(\bm{\xi})\right] \frac{\partial H_{c}}{\partial \bm{\xi}}(\bm{\xi})+ \bm{G}_{c}(\bm{\xi}) \bm{u}_{c},  \\
    \bm{y}_{c} & = \bm{G}_c^\top(\bm{\xi}) \frac{\partial H_{c}}{\partial \bm{\xi}}(\bm{\xi}),
\end{aligned}\label{eq.Controller}
\end{equation}
where $\bm{\xi} \in \mathbb{R}^{n_c}, \bm{u}_c\in \mathbb{R}^m, \bm{y}_c\in \mathbb{R}^m$ are the controller state, input, and output, respectively; $\bm{J}_{c}(\bm{\xi})=-\bm{J}_{c}^{\top}(\bm{\xi})$, $\bm{R}_{c}(\bm{\xi})= \bm{R}_{c}^{\top}(\bm{\xi}) \geq 0$ and $\bm{G}_c(\bm{\xi})$ is a full rank matrix. 
The continuously differentiable function $H_c: \mathbb{R}^{n_c} \rightarrow \mathbb{R}$  is the controller Hamiltonian.
By coupling the plant~\eqref{eq.Plant} with the controller~\eqref{eq.Controller} via the standard negative feedback interconnection
\begin{align}
  \label{eq.PlantControllerInterconnection}
\bm{u} =-\bm{y}_{c}+ \bm{v}, \quad \bm{u}_{c} = \bm{y}+ \bm{v}_{c},
\end{align}
where $\bm{v}, \bm{v}_{c}$ are auxiliary input signals that will be specified later, we  obtain the closed-loop system
\begin{align}
  \label{eq.ClosedLoopSystem}
  \begin{aligned}
    \begin{bmatrix}
      \dot{\bm{x}} \\
      \dot{\bm{\xi}}
    \end{bmatrix}= &
    \left( { \bm{J}_{cl}(\bm{x},\bm{\xi})}
      - {\bm{R}_{cl}(\bm{x}, \bm{\xi})}\right)  \begin{bmatrix}
      \frac{\partial H}{\partial \bm{x}}(\bm{x}) \\
      \frac{\partial H_{c}}{\partial \bm{\xi}}(\bm{\xi})
    \end{bmatrix}\\
    &+\begin{bmatrix}
      \bm{G}(\bm{x}) & 0 \\
      0 & \bm{G}_{c}(\bm{\xi})
    \end{bmatrix} \begin{bmatrix}
      \bm{v} \\
      \bm{v}_{c}
    \end{bmatrix}, \\
    \begin{bmatrix}
      \bm{y} \\
      \bm{y}_{c}
    \end{bmatrix}=&\begin{bmatrix}
      \bm{G}^{\top}(\bm{x}) & 0 \\
      0 & \bm{G}_{c}^{\top}(\bm{\xi})
    \end{bmatrix} \begin{bmatrix}
      \frac{\partial H}{\partial \bm{x}}(\bm{x}) \\
      \frac{\partial H_{c}}{\partial \bm{\xi}}(\bm{\xi})
    \end{bmatrix},
  \end{aligned}
\end{align}
with
\begin{align*}
\bm{J}_{cl}(\bm{x},\bm{\xi})&= \begin{bmatrix}
          \bm{J}( \bm{x}) & - \bm{G}(\bm{x}) \bm{G}_{c}^{\top}(\bm{\xi}) \\
          \bm{G}_{c}(\bm{\xi}) \bm{G}^{\top}(\bm{x}) & \bm{J}_{c}(\bm{\xi})
        \end{bmatrix}, \\
{\bm{R}_{cl}(\bm{x}, \bm{\xi})}&=  \begin{bmatrix}
          \bm{R}(\bm{x}) & 0 \\
          0 & \bm{R}_{c}(\bm{\xi})
        \end{bmatrix}.
\end{align*}
System \eqref{eq.ClosedLoopSystem} is also a PHS, with state space $\mathbb{R}^{n+n_c}$, Hamiltonian $H(\bm{x})+H_{c}(\bm{\xi})$, interconnection matrix $\bm{J}_{c l}(\bm{x}, \bm{\xi})$, dissipation matrix $\bm{R}_{c l}(\bm{x}, \bm{\xi})$, inputs $\left(\bm{v}, \bm{v}_{c}\right)$ and outputs $\left(\bm{y}, \bm{y}_{c}\right)$.
For simplicity, we denote the state for the closed-loop system as $\bm{z}$, i.e., $\bm{z}=[\bm{x}, \bm{\xi}]^\top$.

The control design procedure and the corresponding closed-loop stability properties are described in the following Lemma, which is a simplified version of Proposition 7.1.8 in \cite{vanderSchaft2017}.
\begin{lemma}\label{lem.EnergyCasimirControl}
 If one can find  $\bm{J}_c(\bm{\xi}), \bm{R}_c(\bm{\xi}),  \bm{G}_c(\bm{\xi}), H_c(\bm{\xi})$ for the controller~\eqref{eq.Controller}, a Casimir function $C(\bm{z})$ for the closed-loop system~\eqref{eq.ClosedLoopSystem}, a function $\Phi: \mathbb{R}^2 \rightarrow \mathbb{R}$, and a $\bm{\xi}^*\in \mathbb{R}^{n_c}$, such that the Lyapunov function defined by $V=\Phi(H+H_c, C)$ has a local minimum at $\bm{z}^*=(\bm{x}^*, \bm{\xi}^*)$, i.e.,
  \begin{align}
    \label{eq.LocalMinimumCondition}
    \frac{\partial V}{\partial \bm{z}} \vert_{\bm{z}^*}=0 , \quad
    \frac{\partial^2 V}{\partial \bm{z}^2} \vert_{\bm{z}^*}> 0.
  \end{align}
Then the auxiliary inputs
$$
\bm{v}=-\bm{D} \bm{G}^{\top}(\bm{x}) \frac{\partial V}{\partial \bm{x}}(\bm{x}, \bm{\xi}), \quad \bm{v}_{c}=-\bm{D}_{c} \bm{G}_{c}^{\top}(\bm{\xi}) \frac{\partial V}{\partial \bm{\xi}}(\bm{x}, \bm{\xi})
$$
with $\bm{D}=\bm{D}^{\top} > 0, \bm{D}_{c}=\bm{D}_{c}^{\top} > 0$ asymptotically stabilize~\eqref{eq.ClosedLoopSystem} to $\left(\bm{x}^{*}, \bm{\xi}^{*}\right)$.
\end{lemma}

The main obstacle to the application of the energy Casimir method is that there is no systematic approaches to design parameters and functions appearing in Lemma~\ref{lem.EnergyCasimirControl}. 

\section{Neural Energy Casimir Control\label{sec.NeuralCasimirControl}}
In this section, we describe our method, which uses NNs to approximate the functions in Lemma~\ref{lem.EnergyCasimirControl} and transforms the controller design problem into a parametric optimization problem.

We assume that $\bm{J}_c(\bm{\xi}), \bm{R}_c(\bm{\xi}), \bm{G}_c(\bm{\xi})$ are fixed a prior so that one only need to find candidate functions $H_c,  \Phi, C$ and the controller state $\bm{\xi}^*$ such that~\eqref{eq.LocalMinimumCondition} holds.
We denote with $H_{c, \theta_1}, \Phi_{\theta_2}, C_{\theta_3}$ the NN approximations of the functions $H_c, \Phi$ and $C$, respectively, where $\theta_1, \theta_2, \theta_3$ represent NN parameters.
We have the following optimization targets.
\begin{enumerate}
\item $C_{\theta_3}$ must be a Casimir function for the closed-loop system, i.e., the following condition must be satisfied
  \begin{align}
    \label{eq.CasimirTrainTarget}
    \frac{\partial^{\top} C_{\theta_3}}{\partial \bm{z}}(\bm{z}) \left(\bm{J}_{cl}(\bm{z})- \bm{R}_{cl}(\bm{z})\right)=0,  \forall \bm{z} \in \mathbb{R}^{n+n_c}   \; .
  \end{align}
\item $V_{\theta}$\footnote{Hereafter, we use $\theta$ to denote the parameters of all neural networks.} defined as $V_{\theta}=\Phi_{\theta_2}(H+H_{c, \theta_1}, C_{\theta_3})$ must have a local minimum at $\bm{z}^*=(\bm{x}^*, \bm{\xi}^*)$ for some $\bm{\xi}^*$, i.e., the following condition must be satisfied
  \begin{align}
    \label{eq:1}
  \frac{\partial V_{\theta}}{\partial \bm{z}} \vert_{\bm{z}^*}=0 ,  \quad
 \frac{\partial^2 V_{\theta}}{\partial \bm{z}^2} \vert_{\bm{z}^*}> 0.
  \end{align}
\end{enumerate}

To meet the requirement~\eqref{eq.CasimirTrainTarget}, albeit in an approximate way, one could grid the region of interest, so obtaining a set of points $\Omega=\{\bm{z}_i\}_{i=1}^{N_z}$, and require~\eqref{eq.CasimirTrainTarget} holds on all grid points.
To satisfy~\eqref{eq:1}, one could minimize the norm of the gradient $\frac{\partial V_{\theta}}{\partial \bm{z}} \vert_{\bm{z}^*}$ and penalize the case that the smallest eigenvalue of the Hessian $\frac{\partial^2 V_{\theta}}{\partial \bm{z}^2} \vert_{\bm{z}^*}$ is negative.
In order to consider both goals simultaneously, we propose to the train NNs to minimize the following cost
\begin{multline}
  \label{eq.NonParameterizedCostMinimization}
 \sum_{\bm{z}_i\in \Omega}\|\frac{\partial^\top C_{\theta_3}}{\partial \bm{z}}\vert_{\bm{z}_i} \left(\bm{J}_{cl}(\bm{z}_i)- \bm{R}_{cl}(\bm{z}_i) \right)\| \\ + \|\frac{\partial V_{\theta}}{\partial \bm{z}}\vert_{\bm{z}^*} \|+\text{ReLU}\left(-\lambda_{\min}\left(\frac{\partial^2 V_{\theta}(\bm{z})}{\partial \bm{z}^2}|_{\bm{z}^*}\right) \right),
\end{multline}
where the $\text{ReLU}$ function is defined as  $\text{ReLU}(x)=\max(0,x)$.

The above cost function involves the Jacobian and Hessian of NN outputs w.r.t.\ some inputs. 
Existing machine learning frameworks, such as \textit{pytorch}, provide functions to calculate these quantities, thus allowing to evaluate and optimize the cost~\eqref{eq.NonParameterizedCostMinimization}.

However, since we need to grid the region of interest and minimize the terms  $\|\frac{\partial^\top C_{\theta_3}}{\partial \bm{z}} \left( \bm{J}_{cl}- \bm{R}_{cl}\right) \|$ on all grid points, the computational requirement for training might be high.
Moreover, only approximate Casimirs will be learned and the closed-loop stability cannot be rigorously guaranteed.
In the following, we propose methods to avoid these issues.

\subsection{Parameterization of Casimir Functions}

In this section, we introduce a parameterization of Casimir functions for satisfying the constraints~\eqref{eq.CasimirCondition} by design.
As a result, the first term in the cost~\eqref{eq.NonParameterizedCostMinimization} can be removed, which simplifies the training.

We first assume that $\bm{J}_{cl}$ and $\bm{R}_{cl}$ are constant matrices, and then discuss the more general case where $\bm{J}_{cl}$ and $\bm{R}_{cl}$ are functions of the state $\bm{z}$.
Suppose the dimension of $\ker{\bm{J}_{cl}} \cap \ker{\bm{R}_{cl}} $ is $r$, and let $\bm{v}_1, \ldots, \bm{v}_r$ be the basis of $\ker{\bm{J}}\cap \ker{\bm{R}_{cl}}$.
Then from~\eqref{eq.CasimirCondition}, we have
\begin{align}
  \label{eq.2}
  \frac{\partial C}{\partial \bm{z}}(\bm{z})=\sum_{i=1}^r \alpha_i(\bm{z}) \bm{v}_i
\end{align}
for some functions $\alpha_i(\bm{z}): \mathbb{R}^{n+n_c} \rightarrow \mathbb{R}, \, i=1, \ldots, r$.
A candidate function $C$ satisfying~\eqref{eq.2} is
\begin{align}
  \label{eq.casimir_parameterization}
  C(\bm{z})=K(\sum_{i=1}^r \beta_i(\bm{z}^\top \bm{v}_i))
\end{align}
for some continuously differentiable functions $K: \mathbb{R}\rightarrow \mathbb{R},\, \beta_i: \mathbb{R}\rightarrow \mathbb{R}, \, i=1, \ldots, r$.
Indeed, in this case, $\frac{\partial C}{\partial \bm{z}}$ verifies
\begin{align*}
  \frac{\partial C}{\partial \bm{z}}=\frac{\partial K}{\partial \mathrm{input}} \sum_{i=1}^r \frac{\partial \beta_i}{\partial \mathrm{input}} \bm{v}_i=\sum_{i=1}^r \alpha_i(\bm{z}) \bm{v}_i,
\end{align*}
where $\alpha_i(\bm{z})=\frac{\partial K}{\partial \text{input}}\frac{\partial \beta_i}{\partial \text{input}}$, and $\frac{\partial K}{\partial \text{input}}, \frac{\partial \beta_i}{\partial \text{input}}$ represent the gradient of $K(\cdot)$ and $\beta_i(\cdot)$ with respect to their input, respectively.
The expression~\eqref{eq.casimir_parameterization} provides a parameterization of Casimirs.
Indeed, for any given scalar functions $K(\cdot)$ and $\beta_i(\cdot)$, the RHS of~\eqref{eq.casimir_parameterization} is a Casimir.
Therefore, we can use NNs to represent $K(\cdot)$ and $\beta_i(\cdot)$ and thus further obtaining Casimirs through~\eqref{eq.casimir_parameterization}.

    If $\bm{J}_{cl}$ and $\bm{R}_{cl}$ depends on $\bm{z}$, i.e., in the form of $\bm{J}_{cl}(\bm{z}), \bm{R}_{cl}(\bm{z})$, then from~\eqref{eq.CasimirCondition}, we have
    \begin{align}
      \label{eq.CasimirConditionFunctionCase}
       \frac{\partial C}{\partial \bm{z}}(\bm{z})=\sum_{i=1}^{r(\bm{z})} \alpha_i(\bm{z}) \bm{v}_i(\bm{z}),
    \end{align}
    where $\bm{v}_i(\bm{z}), i=1, \ldots, r(\bm{z})$ form a basis of $\ker \bm{J}_{cl}(\bm{z}) \cap \ker \bm{R}_{cl}(\bm{z})$.
    Denote the RHS of~\eqref{eq.CasimirConditionFunctionCase} as $\bm{F}(\bm{z})$.
    For the existence of $C$ satisfying~\eqref{eq.CasimirConditionFunctionCase}, $\bm{F}(\bm{z})$ must satisfy the integrability condition~\cite[Equation 11.21]{RN11710}
    \begin{align}
      \label{eq:2}
      \frac{\partial F_i}{\partial z_j}(\bm{z})=\frac{\partial F_j}{\partial z_i}(\bm{z}), \quad \forall i, j, \bm{z}
    \end{align}
    where $F_i$ and $z_i$ are the $i$-th element of $\bm{F}$ and $\bm{z}$, respectively.
    The requirement~\eqref{eq:2} is the main difficulty in providing a universal parameterization of general Casimirs.
    Hereafter, we restrict our attention on the case where  $\bm{F}(\bm{z})=[F_{1}(z_1), \ldots, F_{n+n_c}(z_{n+n_c})]^{\top}$, that is the $i$-th element of $\bm{F}$ is only a function of the $i$-th element of $\bm{z}$.
    A candidate $C$ satisfying~\eqref{eq.CasimirConditionFunctionCase} is then given by
    \begin{align*}
     C=K(\int F_{1}(z_1)dz_1+\ldots +\int F_{n+n_c}(z_{n+n_c})dz_{n+n_c}),
    \end{align*}
    for some continuously differentiable function $K: \mathbb{R} \rightarrow \mathbb{R}$.

\subsection{Neural Energy Casimir Control with Parameterized Casimir Functions}

With the proposed parameterization of Casimir functions, we modify the neural energy Casimir control design as follows, where we assume $\bm{J}_{cl}, \bm{R}_{cl}$ are constant matrices\footnote{When  $\bm{J}_{cl}, \bm{R}_{cl}$ are functions of $\bm{z}$, and if we can find a parameterization of $C$, the modified neural energy Casimir control can be obtained similarly.}.
We use the NNs $H_{c, \theta_{1}}(), \Phi_{\theta_{2}}()$ to approximate $H_c()$ and $\Phi()$ in Lemma~\ref{lem.EnergyCasimirControl}, respectively, and use the NNs $K_{\theta_{3}}()$ and $\beta_{i, \theta_{i+3}}()$ to approximate $K()$ and $\beta_i(), , i=1, \ldots,r$ in~\eqref{eq.casimir_parameterization}, respectively, where
$\theta_i, i=1, \ldots, r+3$ are NN parameters.
Then we define the neural Lyapunov function
\begin{align*}
  V_{\theta}(\bm{z})= \Phi_{\theta_{2}}(H(\bm{x})+ H_{c, \theta_{1}}(\bm{\xi}),  K_{\theta_{3}}(\sum_{i=1}^r \beta_{i, \theta_{i+3}}(\bm{z}^\top \bm{v}_i)))
\end{align*}
and solve the following optimization problem
\begin{multline}
\min_{\theta_1, \ldots, \theta_{r+3}, \bm{\xi}^*} \| \frac{\partial V_{\theta}(\bm{z})}{\partial \bm{z}} \vert_{\bm{z}^*} \| \\ + \text{ReLU}\left(-\lambda_{\min}\left(\frac{\partial^2 V_{\theta}(\bm{z})}{\partial \bm{z}^2}|_{\bm{z}^*}-a \bm{I}\right) \right),  \label{eq.Loss}
\end{multline}
where the term $a\bm{I}$ with $a>0$ is added to the Hessian regularization to promote local convexity of $V_{\theta}$ around the closed-loop equilibrium.
This can improve the control performance as shown in the next section.
\begin{remark}
  The proposed method is computationally less expensive than the NN-based control design with stability guarantees in~\cite{chang2020neural, dai2021lyapunov} for general nonlinear systems.
  In~\cite{chang2020neural, dai2021lyapunov}, a region of interest is firstly sampled, then the NN controller is trained on all sampled points.
  Moreover, a verification step is employed to check whether the learned candidate Lyapunov function is a Lyapunov function.
  The above training and verification procedure requires considerable computations.
  Instead, the proposed training objective~\eqref{eq.Loss} requires to solve a simpler optimization problem and therefore is easier to implement.
\end{remark}

\subsection{Performance Analysis}

In this section, we analyze the performance of the neural energy Casimir control method.
We show that when the loss~\eqref{eq.Loss} is sufficiently small after training, the distance between the equilibrium point of the learned Lyapunov function and the desired equilibrium point $\bm{z}^*$ is also small.
Suppose after training, we obtain a small loss $\epsilon$
\begin{align}
  \label{eq.FinalTrainLoss}
\epsilon= \| \frac{\partial V_{\theta}(\bm{z})}{\partial \bm{z}} \vert_{\bm{z}^*} \|+ \text{ReLU}\left(-\lambda_{\min}\left(\frac{\partial^2 V_{\theta}(\bm{z})}{\partial \bm{z}^2}|_{\bm{z}^*}-a I\right) \right).
\end{align}
From~\eqref{eq.FinalTrainLoss}, we have
\begin{gather}
  \| \frac{\partial V_{\theta}(\bm{z})}{\partial \bm{z}} \vert_{\bm{z}^*} \|\le \epsilon,\label{eq.LossGradTermUpperBound}\\
  \text{ReLU}\left(-\lambda_{\min}\left(\frac{\partial^2 V_{\theta}(\bm{z})}{\partial \bm{z}^2}|_{\bm{z}^*}-a I\right) \right)\le \epsilon. \label{eq.LossHessianTermUpperBound}
\end{gather}
Since $\epsilon$ is close to zero and $a$ is larger than zero, we have that $a-\epsilon>0$.
Then from~\eqref{eq.LossHessianTermUpperBound} we have
$  \frac{\partial^2 V_{\theta}(\bm{z})}{\partial \bm{z}^2}|_{\bm{z}^*} \ge (a-\epsilon) I$.
By left and right multiplying the above inequality with $\left( \frac{\partial^2 V_{\theta}(\bm{z})}{\partial \bm{z}^2}|_{\bm{z}^*} \right)^{-\frac12}$, we have $\left( \frac{\partial^2 V_{\theta}(\bm{z})}{\partial \bm{z}^2}|_{\bm{z}^*}\right)^{-1}\le \frac{1}{a-\epsilon} I$, therefore
\begin{align}
\| \left(\frac{\partial^2 V_{\theta}(\bm{z})}{\partial \bm{z}^2}|_{\bm{z}^*}\right)^{-1}\| \le 1/(a-\epsilon),\label{eq:HessianInverseBound}
\end{align}
where the matrix norm is the spectral norm.

The gradient $\frac{\partial V_{\theta}(\bm{z})}{\partial \bm{z}} \vert_{\bm{z}}$ can be expanded using Taylor's theorem as
\begin{align}
\label{eq:3}
  \frac{\partial V_{\theta}(\bm{z})}{\partial \bm{z}} \vert_{\bm{z}}=  \frac{\partial V_{\theta}(\bm{z})}{\partial \bm{z}} \vert_{\bm{z}^*}+ \frac{\partial^2 V_{\theta}(\bm{z})}{\partial \bm{z}^2}|_{\bm{z}^*}(\bm{z}-\bm{z}^*)+o(\bm{z}-\bm{z}^*)
\end{align}
If $\epsilon$ is sufficiently small,  from~\eqref{eq.LossGradTermUpperBound}, $\frac{\partial V_{\theta}(\bm{z})}{\partial \bm{z}} \vert_{\bm{z}^*}$ is also close to zero.
Since $\frac{\partial^2 V_{\theta}(\bm{z})}{\partial \bm{z}^2}|_{\bm{z}^*}>0$, then in a sufficiently small neighborhood of $\bm{z}^*$, there exists $\bar{\bm{z}}$, such that  $\frac{\partial V_{\theta}(\bm{z})}{\partial\bm{z}}\vert_{\bar{\bm{z}}}=0$.
Moreover, the continuity of $\frac{\partial^2 V_{\theta}(\bm{z})}{\partial \bm{z}^2}|_{\bm{z}^*}$ implies that $\frac{\partial^2 V_{\theta}(\bm{z})}{\partial \bm{z}^2}|_{\bar{\bm{z}}}>0$, which means $\bar{\bm{z}}$ is a local minimum of $V_\theta$ in a small neighborhood of $\bm{z}^*$.
From~\eqref{eq:3}, we also have
\begin{align*}
  \|\bar{\bm{z}}-\bm{z}^*\| \le \| \frac{\partial^2 V_{\theta}(\bm{z})}{\partial \bm{z}^2}|_{\bm{z}^*}^{-1}\| \left( \|\frac{\partial V_{\theta}(\bm{z})}{\partial \bm{z}} \vert_{\bm{z}^*} \| + \|o(\bm{z}-\bm{z}^*) \| \right),
\end{align*}
Omitting the high-order terms and in view of~\eqref{eq:HessianInverseBound}, we can obtain
\begin{align}
  \label{eq.ErrorUpperBound}
\|\bar{\bm{z}}-\bm{z}^*\| \le \frac{\epsilon}{a-\epsilon}.  
\end{align}
Since $\bar{\bm{z}}$ is the local minimum of $V_\theta$, when applying the designed neural energy Casimir controller, the closed-loop system state would converge to $\bar{\bm{z}}$.
Therefore, the result~\eqref{eq.ErrorUpperBound} implies that if the cost~\eqref{eq.Loss} to close to zero after minimization, we can effectively reduce the set-point control bias caused by the NNs approximation errors.

\begin{remark}
  The (neural) energy Casimir method can only guarantee the local stability of the desired equilibrium, since it only requires the desired equilibrium to be a local minimum of the Lyapunov function.
  However, from the simulation example given in the next section, we can observe that the learned Lyapunov function can have a large region of attraction.
  One could tune the region of attraction achieved with the neural energy Casimir method by including additional regularization terms in the cost~\eqref{eq.Loss}.
  For example, similar to~\cite{chang2020neural}, we can sample the region of interest to obtain data points $\{\bm{z}_1, \ldots, \bm{z}_{N}\}$, and add the regularization term $\text{ReLU}(\frac{1}{N}\sum_{i=1}^N (\|\bm{z}_i\|^2-\gamma V_{\theta}(\bm{z}_i)))$ to the cost~\eqref{eq.Loss}, where $\gamma>0$ is a hyperparameter, which regulates how fast the Lyapunov function value increases with respect to its input.
  Moreover, after training, similar to~\cite{chang2020neural}, one can use satisfiability modulo theory solvers to verify whether the learned $V_\theta(\bm{z})$ is a Lyapunov function over all state vectors in the region of interest.
  \end{remark}
  
\section{Simulations\label{sec.Simulation}}
In this section, we illustrate the performance of the proposed control design method via simulations.
We consider the set-point control of the pendulum system described in~\cite[Example 7.1.12]{vanderSchaft2017} and given by
\begin{align}
  \label{eq.PendulumPlant}
  \begin{bmatrix}
    \dot{q}\\
    \dot{p}
  \end{bmatrix}
  =
  \begin{bmatrix}
    0 & 1\\
    -1 & 0
  \end{bmatrix}
         \begin{bmatrix}
           \sin(q)\\
           p
         \end{bmatrix}
  +
  \begin{bmatrix}
    0 \\
    1
  \end{bmatrix} u, \quad
  y=p,
\end{align}
where $q$ is the angle and $p$ is the momentum, and $u$ is the input torque.
The system~\eqref{eq.PendulumPlant} is in the PHS form~\eqref{eq.Plant} with

\begin{align*}
&H=\frac12 p^2+(1-\cos(q)),\quad \bm{J}=
\begin{bmatrix}
  0 & 1\\
  -1 &0
\end{bmatrix},\\
       &\bm{R}=
\begin{bmatrix}
  0 & 0 \\
  0 & 0
\end{bmatrix},
\quad \bm{G}=
\begin{bmatrix}
  0\\1
\end{bmatrix}.
\end{align*}
Suppose we want to stabilize the pendulum at the non-zero angle $q^*=\frac{\pi}{4}$ with $p^*=0$. We consider the controller
\begin{align*}
  \dot{\xi}=y+v_c, \quad u=-\frac{\partial H_c}{\partial \xi}(\xi) +v,
\end{align*}
which corresponds to~\eqref{eq.Controller} with $\bm{J}_c=0, \bm{R}_c=0, \bm{G}_c=1$, interconnected with the plant~\eqref{eq.PendulumPlant} as in~\eqref{eq.PlantControllerInterconnection}.
Therefore, the closed-loop system is in the PHS form~\eqref{eq.ClosedLoopSystem} with 
\begin{align*}
\bm{J}_{cl}=  \begin{bmatrix}
    0 & 1 & 0 \\
  -1 & 0 & -1\\
  0 & 1 & 0
\end{bmatrix}, \quad
          \bm{R}_{cl}=
          \begin{bmatrix}
    0 & 0 & 0 \\
  0 & 0 & 0\\
  0 & 0 & 0
\end{bmatrix}.
\end{align*}
Since $\bm{J}_{cl}$ has a $1$-dimensional kernel space spanned by the vector $[1,0,-1]^\top$, from~\eqref{eq.casimir_parameterization}, the Casimir function can be parameterized as $C=K(q-\xi)$ for some function $K: \mathbb{R}\rightarrow \mathbb{R}$.
We could use a NN to approximate the function $\Phi()$ and optimize its parameters.
Alternatively, we can also manually choose the form of $\Phi()$, which can improve training speed.
In this simulation, we select $\Phi=H+H_c+C$.

\begin{figure}
    \centering
    \includegraphics[width=0.5\textwidth]{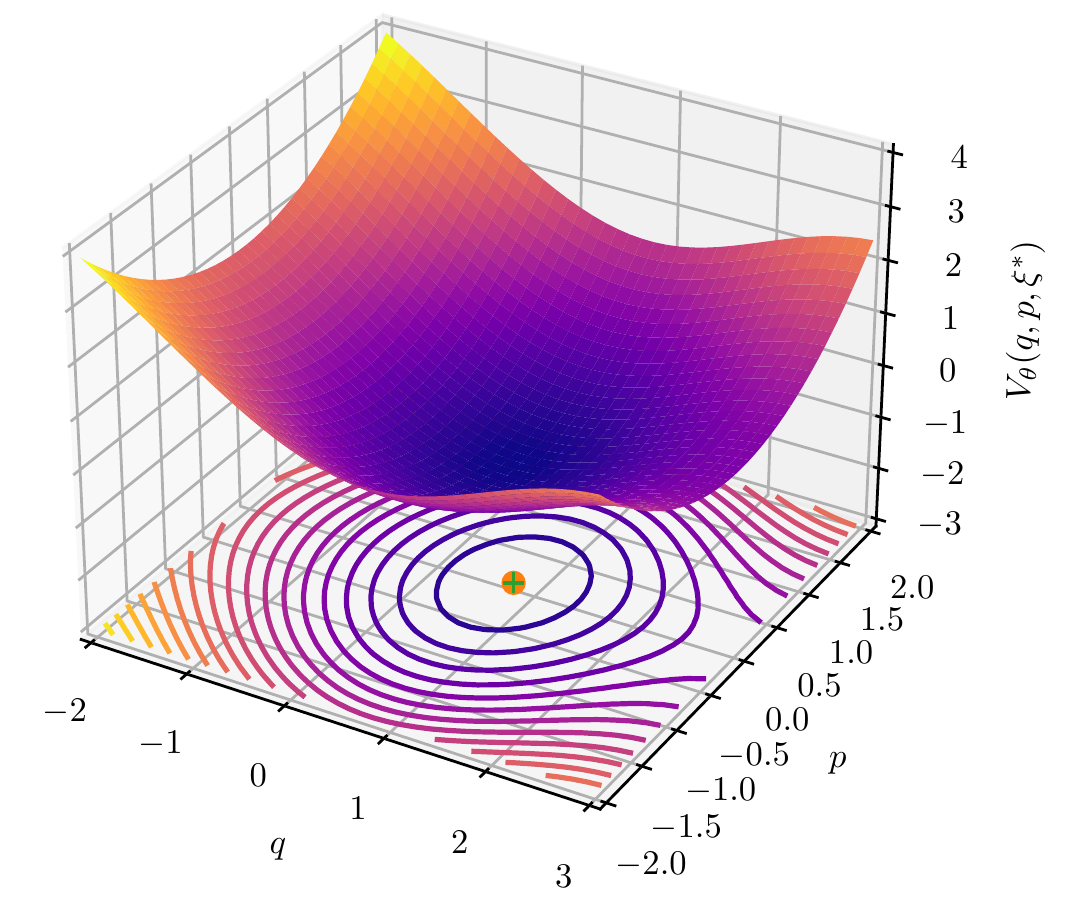}
    \caption{Learned Lyapunov function $V_{\theta}(q, p, \xi^*)$, where \orangecircle denotes the desired equilibrium and ${\color{OliveGreen}\textbf{\texttt{+}}}$ is the projected minimum point of the learned Lyapunov function}
        \label{fig.PendulumLyapunovPlot}
\end{figure}

\begin{figure}
    \centering
    \includegraphics[width=0.5\textwidth]{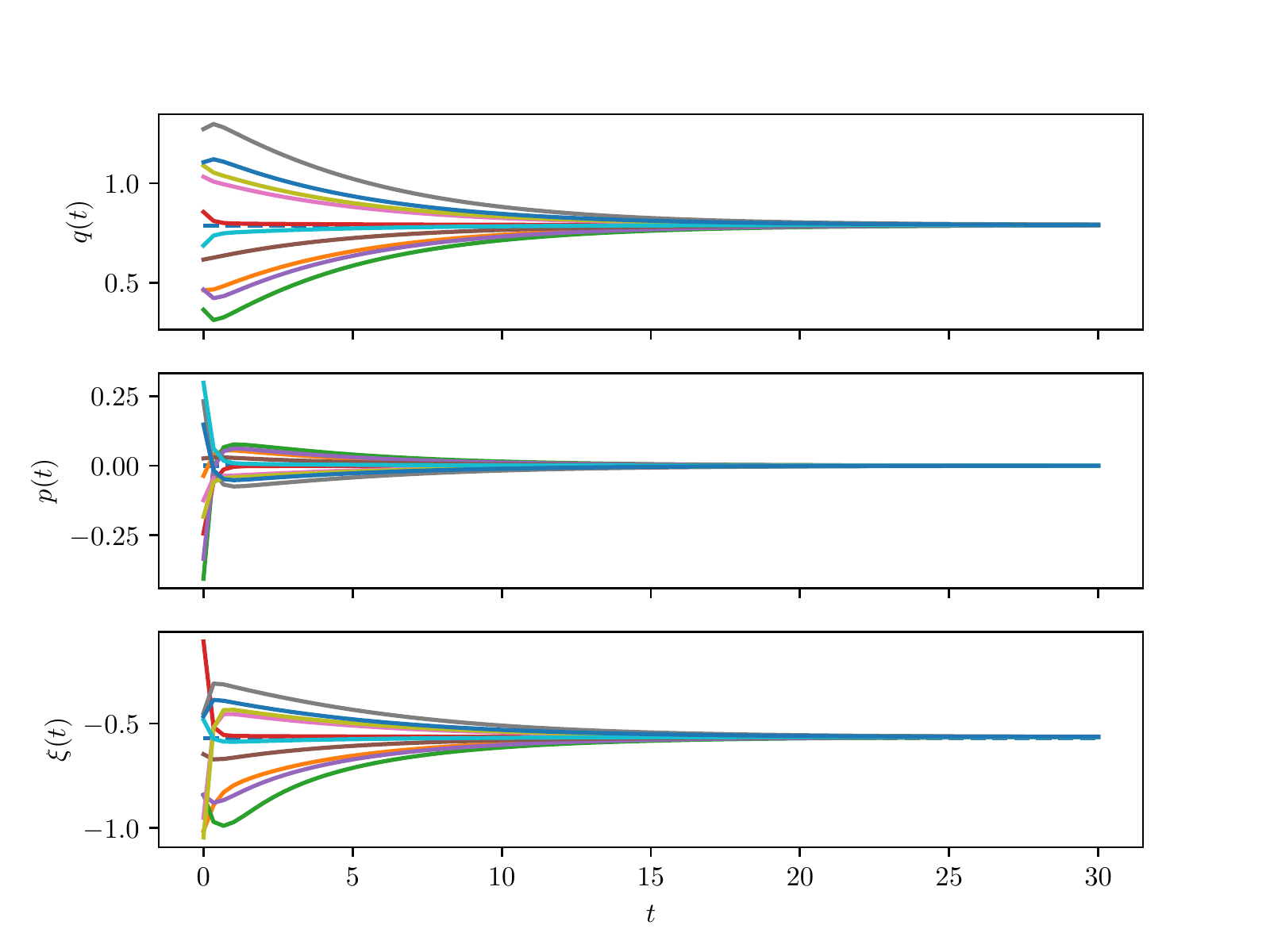}
    \caption{Closed-loop response under neural Casimir controller, where the dashed line represents the desired equilibrium}
        \label{fig.PendulumResponsePlot}
\end{figure}

Under these settings, traditional energy Casimir control method requires to design the functions $K(\cdot)$ and $H_c(\cdot)$ and $\bm{\xi}^*$, such that the following conditions holds
\begin{gather*}
\sin q^{*}+\frac{\partial K}{\partial \text{input}}\vert_{q^{*}-\xi^{*}}=0, \\
-\frac{\partial K}{\partial \text{input}} \vert_{q^{*}-\xi^{*}}+\frac{\partial H_{c}}{\partial \xi}\vert_{\xi^{*}}=0,\\
        \begin{bmatrix}
          \cos q^{*}+\frac{\partial^{2} K}{\partial \text{input}^{2}}\vert_{q^{*}-\xi^{*}} & 0 & -\frac{\partial^{2} K}{\partial \text{input}^{2}}\vert_{q^{*}-\xi^{*}} \\
          0 & 1 & 0 \\
          -\frac{\partial^{2} K}{\partial \text{input}^{2}}\vert_{q^{*}-\xi^{*}} & 0 & \frac{\partial^{2} K}{\partial \text{input}^{2}}\vert_{q^{*}-\xi^{*}}+\frac{\partial^{2} H_{c}}{\partial \xi^{2}}\vert_{\xi^{*}}
        \end{bmatrix}>0,
      \end{gather*}
      which are not straightforward to satisfy.
     Instead, in neural energy Casimir control, we can use NNs to approximate $K$ and $H_c$ and optimize the network parameters and $\bm{\xi}^*$ to fulfill the above constraints.
      We use multilayer perception (MLP) NNs to approximate the function $K$ and the controller Hamiltonian $H_c$.
      Specifically, we use MLP NNs with $1$ input layer, $1$ hidden layer, and $1$ output layer, where the width of the hidden layer is $64$, to approximate $K$, and $32$, to approximate $H_c$.
      As for optimization, we use the ADAM algorithm~\cite{kingma2014adam} with step size $1\times 10^{-4}$.
We then train the NNs by optimizing the cost~\eqref{eq.Loss}, where $a=0.5$.
The control gains are set to $\bm{D}=5$ and $\bm{D}_c=6$. 
We highlight that, the NN structures and the optimizer step size have been chosen by trial and error.
We gradually increase the depth and width of the NNs, and decrease the optimizer step size, until we have obtained a sufficiently small training loss~\footnote{The code for the simulation is built on pytorch-lightning and can be accessed at: \url{https://github.com/DecodEPFL/neural_energy_casimir_control}}.
After training for $2000$ epochs, we let $\xi=\xi^*$ and obtain the learned Lyapunov function in the $(q, p)$ space, shown in Fig.~\ref{fig.PendulumLyapunovPlot}.
We can observe that it achieves the minimum around the desired equilibrium point $q=\frac{\pi}{4}, p=0$.
Moreover, it is convex when $(q, p)\in \mathcal{A}= [-2, 2]\times[-2,2]$, which implies the region of attraction is at least $\mathcal{A}$ and therefore quite large.
The closed-loop response for $10$ randomly generated  initial states are shown in Fig.~\ref{fig.PendulumResponsePlot}.
One can notice that the trajectories converge to the desired equilibrium, hence demonstrating the effectiveness of the proposed control method.

\begin{figure}
    \centering
    \includegraphics[width=0.5\textwidth]{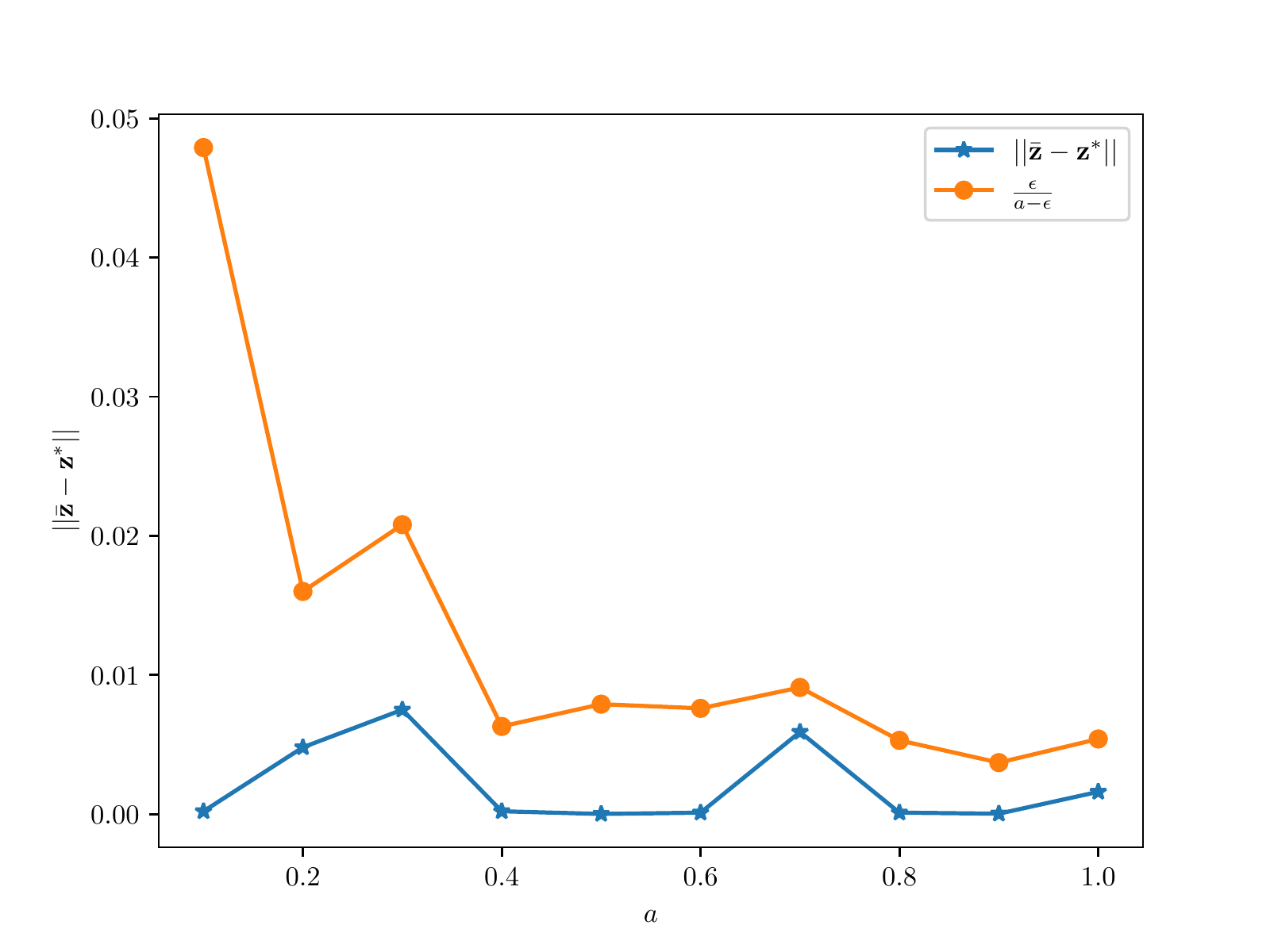}
    \caption{The error $\|\bar{\bm{z}}-\bm{z}^*\|$ and the bound $\frac{\epsilon}{a - \epsilon}$ with respect to different values of $a$}
    \label{fig:upperBound}
\end{figure}

After training, we obtain the loss $\epsilon= 0.0050$ and the controller equilibrium $\xi^*=-0.5693$.
Therefore, the upper bound in \eqref{eq.ErrorUpperBound} is $\frac{\epsilon}{a-\epsilon}=0.0101$.
Moreover, we randomly select $1$ trajectory from the $10$ generated trajectories, and find that  the trajectory converges to  $([ 7.9093\times 10^{-1},  -2.2693 \times 10^{-4}, -5.6329 \times 10^{-1}])$, which is also the minimum $\bm{\bar{z}}$ of the learned Lyapunov function.
Since $\bm{z}^*=[ 0.7854,  0.0000, -0.5693]$, we have $\|\bar{\bm{z}}-\bm{z}^*\|=0.0082$.
As a result,  the bound $\|\bar{\bm{z}}-\bm{z}^*\|\le \frac{\epsilon}{a-\epsilon}$ in~\eqref{eq.ErrorUpperBound} is verified.
Finally, we train the NNs in the same setting for different values of $a$.
  In Fig. \ref{fig:upperBound}, we plot the term $\frac{\epsilon}{a-\epsilon}$ and the error $\|\bar{\bm{z}}-\bm{z}^*\|$ as a function of $a$. We can observe that the bound \eqref{eq.ErrorUpperBound} is verified in all cases.


\section{Conclusions\label{sec.Conclusions}}
We propose a NN-based approach to facilitate the control of PHS through the energy Casimir method.
Our approach does not require solutions of convoluted PDEs for  equilibrium assignment.
Instead, for achieving closed-loop stability, we use parameterizations of Casimir functions and provide regularization terms penalizing the Jacobian and Hessian of the neural Lyapunov function at the desired equilibrium point.
Moreover, we prove that the difference between the desired and achieved equilibrium point can be bounded in terms of the training loss.
Further work will be devoted to analyze the performance of different classes of deep neural networks (DNNs) including Hamiltonian-DNNs \cite{galimberti2021hamiltonian} which match the Hamiltonian structure of the system and controller.
Moreover, we aim at extending the proposed framework to other controller design procedures for PHSs.

\bibliographystyle{IEEEtran}
\bibliography{references}
\end{document}